# A high aspect ratio Fin-Ion Sensitive Field Effect Transistor: compromises towards better electrochemical bio-sensing

Serena Rollo[a,b], Dipti Rani[a], Renaud Leturcq[a], Wouter Olthuis[b], César Pascual García[a†]

a. Materials Research and Technology Department, Luxembourg Institute of Science and Technology (LIST), Belvaux, Luxembourg
b. BIOS Lab on Chip Group, MESA+ Institute for Nanotechnology, University of Twente, Enschede, The Netherlands.

† Corresponding author: cesar.pascual@list.lu .

The development of next generation medicines demand more sensitive and reliable label free sensing able to cope with increasing needs of multiplexing and shorter times to results. Field effect transistor-based biosensors emerge as one of the main possible technologies to cover the existing gap. The general trend for the sensors has been miniaturisation with the expectation of improving sensitivity and response time, but presenting issues with reproducibility and noise level. Here we propose a Fin-Field Effect Transistor (FinFET) with a high heigth to width aspect ratio for electrochemical biosensing solving the issue of nanosensors in terms of reproducibility and noise, while keeping the fast response time. We fabricated different devices and characterised their performance with their response to the pH changes that fitted to a Nernst-Poisson model. The experimental data were compared with simulations of devices with different aspect ratio, stablishing an advantage in total signal and linearity for the FinFETs with higher aspect ratio. In addition, these FinFETs promise the optimisation of reliability and efficiency in terms of limits of detection, for which the interplay of the size and geometry of the sensor with the diffusion of the analytes plays a pivotal role.

## Introduction

Silicon nanowire-field effect transistors (SiNW-FETs) are one of the candidates to be among the building blocks of the next generation molecular diagnostic devices as they offer label-free detection, are miniaturized and thus can be integrated on a microfluidic platform for rapid and low-cost assay [1-3]. Their three-dimensional configuration makes them more efficient than planar FETs to detect ultra-low concentrations of analytes due to a better gating effect [4-6]. There is a second advantage linked to their geometry. The species contributing to the sensing signal are detected after binding to the functionalized surface of the device depleting the medium, which creates a concentration gradient. In order to reach the equilibrium signal the analytes must diffuse and bind to the surface of the sensor. At low concentrations, the diffusion of the molecules towards the sensor requires longer incubation times to reach the limits of detection that provide a readable signal. In a planar sensor, like a traditional ISFET, the remaining analytes can only diffuse in one dimension perpendicular to its surface. In SiNWs the diffusion occurs in the two dimensions perpendicular to the wire (hemicylindrical), which results in a much faster adsorption of analytes[7-9]. SiNWs have shown low limits of detection for different biomarkers comprising DNA [10-14] and proteins [2, 15-17] in different media including biological fluids [18-21] and tissues [22, 23]. Owing to the well-known nanofabrication methods, and low operational power, Si NW-FETs can be easily integrated into CMOS chips where transducers and necessary circuits for signal processing are integrated on a single chip [2, 10, 15, 24]. Such devices bring promises in order to have cost effective point of care (POC) and highly multiplexed sensors for personalised precision medicines [25].

The transducing response of SiNW-FET is faster than the diffusion and binding, which would qualify them for real-time sensing. However, the main limitation for real-time measurement is the Debye screening, which causes a limitation of the sensing region due to the counter-ions present and shielding the species at high ionic strength electrolytes [26, 27]. For this reason, many of the experiments reported in literature do not measure in physiological conditions, and consist first of an incubation step followed by washing and measurement at lower ionic strengths, which results in an end-point result detrimental for the biological potential of SiNW's because they cannot evaluate kinematic constants like the molecular affinity [28-30].

Regardless of all their perspectives and the intense research activities during the last decades, SiNW-FET still have not been introduced into any clinical application, mainly due to their problems of repeatability and reliability. Due to their small size, they are intrinsically sensitive to the fabrication defects and the more difficult to control functionalisation of bio-receptors that bring fluctuations in the transport that in the case of NWs propagate along the conduction channel. Current approaches are directed to improve the fabrication methods and the material reliability from the quality control on larger production batches [5, 31] and the control of composition homogeneity of the dielectric sensitive layer [32-34]. However, the current trend is the miniaturisation of NWs to exploit the advantages of small size and the three dimensional geometry, disregarding the difficulties to achieve reliable homogeneous functionalisation in single devices with few tenths of nanometres. Due to the small cross section of the conductive channels, SiNWs carry relatively small currents, which makes their integration also more difficult due to the required voltage necessary to polarise high resistive devices. The requirements of more accurate instrumentation in small devices, together with their lack of reliability increases their production costs that cannot cope with the required quality control that would be necessary. NW arrays measured in parallel can increase the total signal, and mitigate some of the lack of reliability due to the averaged device variability, but this increases the overall device footprint, which limits their efficiency at low analyte concentrations in diffusion-limited processes [7-9,35] and jeopardizes the viability of massive multiplexing due to the increase of footprint.

In this work we approach the problem from the point of view of the design by combining advantages of nano-sensors with the reliability of planar devices, being able to keep the advantages of three dimensional bio-sensors. We propose a novel Fin configuration for a FET with a high aspect ratio of width ($W$) and height ($h$) of the device. Figure 1 (a) and (b) schematically show the proposed FinFET and the cross section at different ion concentrations, respectively. Setting $W$ close to twice the maximum depletion region ($W_D$) expected during the dynamic range, the physical aspect ratio $h:W$ and the electrostatic parameters specifying the channel doping, the dielectric constant and the thickness of the oxide are related. This relationship between the shape of the device and the electrostatic parameters has deep consequences for the performance of the wire in terms of total signal and linearity. We tested the operation of the FinFET against pH sensitivity. Our p-doped FinFETs work in depletion mode for positive charges like protons.

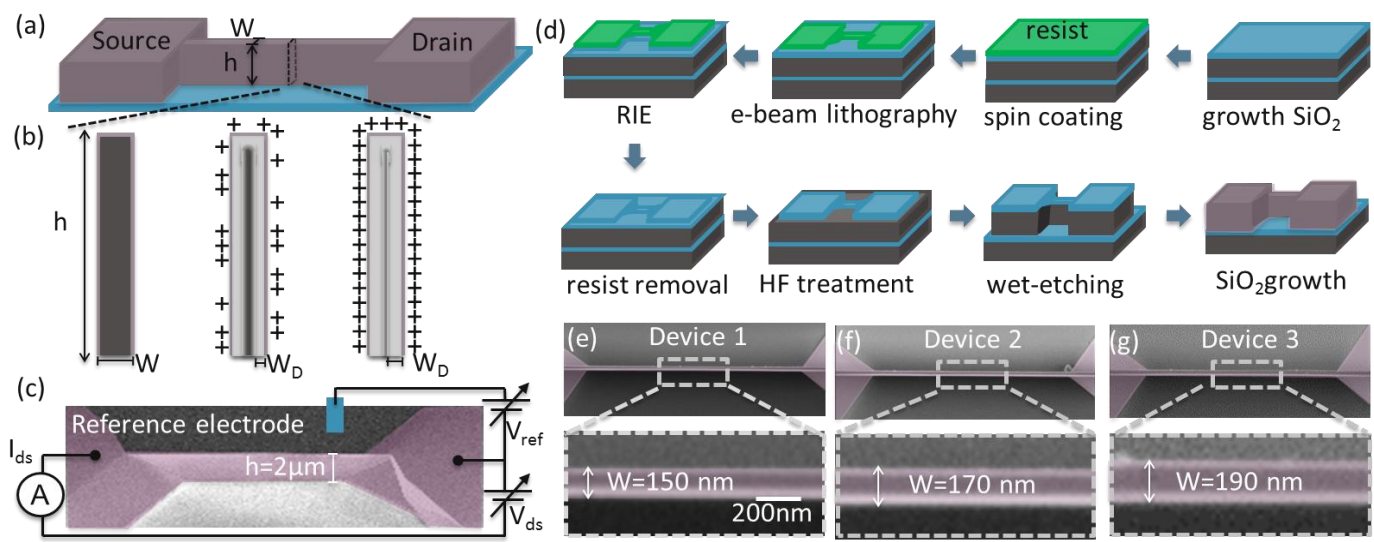

**Figure 1** (a) Schematic representation of the FinFET. (b) Representation of the FINFET from low ion concentration to higher ion concentrations, switching from fully conductive to almost fully depleted. The darker grey represents in this case the region populated with positive carriers, while the lighter grey the depleted region. (c) SEM tilted picture of a representative device with the schematic representation of the measuring set-up. (d) Summary of the fabrication process: 1x1 cm$^2$ diced samples of <110> SOI were used. 50 nm thermal SiO$_2$ was grown on top the SOI substrate. We spin casted a negative resist and exposed the FinFET mask design with e-beam lithography. We transferred the patterns by RIE leaving a few nm of oxide that was later removed by HF. Vertical walls were achieved by anisotropic wet etching of the <111> facets. The mask was removed by HF before contacting the samples. (e) to (g) Top view images of three representative devices reported in the article, with heights of 2.16 µm and top widths 150, 170 and 190 nm respectively, referred as devices 1 to 3 respectively. A zoomed region showing details of the top surface is also shown.

Considering the doping density of the starting SOI substrate, we designed our wires to switch from fully conductive (no depletion) to nearly fully depleted in a pH range of ~ 8 units (represented in fig. 1 (b)). The larger total surface area compared to SiNWs decreases the impact of surface defects. The chosen configuration of FinFET provided a larger cross-section compared to traditional NWs (or NW arrays) responsible for a large output current with an improved linear response. The improvement in the output current would contribute to a higher signal to noise ratio and the 2D conductivity of the vertical Fin would improve the reliability decreasing the sensitivity to local defects. We argue with geometrical reasons linked to the diffusion time that the increased size of our sensor along the height with respect to NW's would not decrease significantly the detection limits. Finally, the increased size in the vertical direction enhances the total surface area of the sensor, which would increase also the reliability towards bio-functionalisation.

## Results and discussions

Figure 1 (c) shows a SEM image of a representative device with the schematic representation of the measurement set up. We fabricated our Si FinFETs by anisotropic wet etching on a p-doped silicon on insulator (SOI) substrate with a 2.2 ± 0.1 µm thick silicon device layer (<110> oriented) with conductivity of

0.115 Ω·cm (equivalent doping $10^{17}$/cm$^3$) and a 1 µm thick buried *SiO$_2$* procured from Ultrasil Corporation. The overall fabrication process is illustrated in fig. 1 (d). Briefly, we lithographed the masked of the wires on a thermally grown thin *SiO$_2$* that was transferred by Reactive Ion etching and then treated with HF to obtain smooth surface. The anisotropic etching was achieved with a 25 % wt Tetramethylammonium hydroxide, 8.5 %vol of Isopropanol water solution using the proper orientation to obtain vertical walls. 20 nm of *SiO$_2$* was grown to have a pH sensitive layer. The ohmic contacts were defined by optical lithography and e-beam evaporation. The devices were contacted with lithographically designed leads and the sample was protected with an epoxy (SU8) layer leaving open the region of the FinFET's. The sample was mounted on a PCB, and wire bonded for measuring. The bonded wires were protected with a medical grade epoxy glue (details of fabrication in SI 1).

The resulting wires had a length of 14 µm at the middle of the FinFET, which we use later as the average length for calculations. We fabricated two samples each consisting of eleven FinFETs all showing coherent data with similar transfer characteristics and consistent output behaviour. Here we report the results from three representative devices from one of the samples where we carried out a more extensive characterisation. The height of the samples was measured 2.16 µm at the source and drain before the ohmic contacts with a KLA Tencor P-17 profilometer. The three considered devices are characterized by top widths (*W*) of 150, 170 and 190 nm, which account for aspect ratio *h:W* of 15, 13 and 12, respectively, herein referred as devices 1 to 3 from the smallest to the biggest (figs. 1 (e) to (g)). After implementing ohmic contacts, metal leads and passivation of the contacts with optical lithography, the samples were measured in an electrolyte consisting of a solution of *KH$_2$PO$_4$*, citric and boric acids at 0.1 M all, mixed with a *KNO$_3$* 0.1 M solution in equal volume proportion, which had a final pH of 2.5. More basic pH were obtained by titration using a 0.1 M solution of *KOH*. With this procedure, the total ionic strength remained constant at 0.1 M (details in SI 2).

The *SiO$_2$* surface of our wires is sensitive to the proton concentration, which changes the potential at the interface ($\Psi_0$) depleting the channel region of positive carriers. The non-depleted region contributes to the conductance with an ohmic contribution so that the total conductance is calculated from the remaining cross section, length and conductivity of the non-depleted region using a Nernst-Poisson model (we present detailed description in supporting information SI 6). In order to find the working range of drain source ($V_{ds}$) and reference electrode ($V_{ref}$) voltages in which the response of the devices can be described in terms of ohmic contribution of the non-depleted region we run an electrical characterization at neutral pH. The drain source currents ($I_{ds}$'s) from the three devices were recorded within $V_{ds}$ between -300 and 300 mV at different fixed values of $V_{ref}$ between -300 and 300 mV separated by steps of 100 mV (details of the electrical characterization and data from a representative device is shown SI 3-4). We observed a linear behaviour of the output characteristics in $V_{ds}$ voltages around the range between -100 and 100 mV at $V_{ref}$ between -200 and 200 mV. We restricted our study to a more conservative

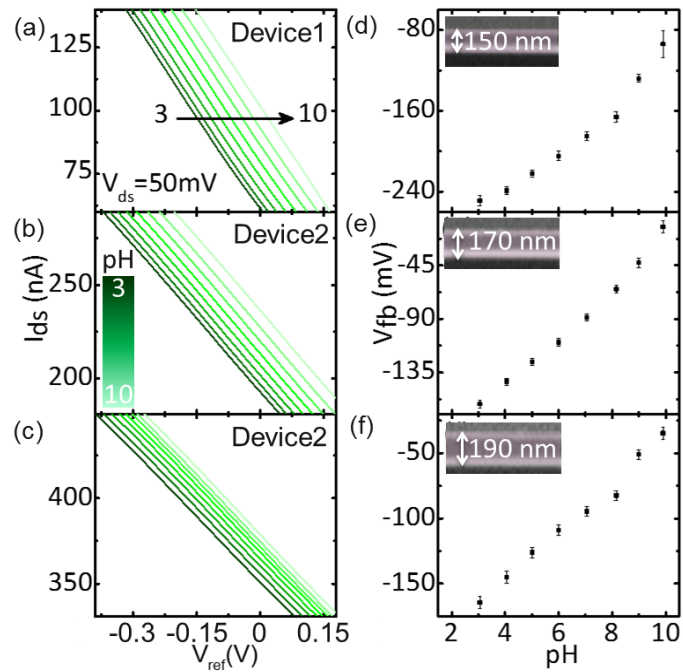

**Figure 2** (a) to (c) are the transfer $I_{ds}$ vs. $V_{ref}$ characteristics at eight different electrolyte pH values from 3 to 10 in steps of 1 represented with the green colour scale shown in (b) for devices 1 to 3, respectively. (d) to (f) are the evaluated flat band voltage potentials of the wires at different pH for the deices 1 to 3, respectively.

working range for $V_{ds}$ between of -50 to 50 mV where our samples regardless the low working potentials, exhibited a good signal to noise ratio. As expected from the Nernst-Poisson model, $I_{ds}$ increased with lower $V_{ref}$ due to the attraction of negative potentials on the positive carriers. From these results we identified the working range within -50 < $V_{ds}$ < 50 mV and -200 < $V_{ref}$ < 200 mV where the electrical transport of the FinFET devices can be described by the ohmic contribution of the non-depleted region using a Nernst-Poisson model.

**A. Transfer characteristics of pH response**

After the neutral pH characterization of the FinFET devices, they were deployed for electrolyte pH sensing. We measured the transfer characteristics sweeping $V_{ref}$ at a fixed $V_{ds}$ (50 mV) and varying the electrolyte pH by manual titration from acidic towards basic values. Figures 2 (a) to (c) show the transfer $I_{ds}$ vs. $V_{ref}$ characteristics at eight different electrolyte pH values from 3 to 10 in steps of 1 represented with the green colour scale shown in (b) for devices 1 to 3, respectively. As expected from the wider cross-section, the devices with more width had higher currents than the smaller ones. The drain current increases with increase in pH of the electrolyte solution, because higher pH induces more negative surface potentials at the silicon oxide sensing layer surface, which in turns creates an attractive electric field for the positive charges inside the channel. In terms that are more standard for ISFETs operation, there is a shift needed of the reference electrode to keep the same depletion region.

The shift of the surface potential $\Psi_0$ was evaluated from the flat band voltage calculating the minimum of the first derivative of the transfer curves at different electrolyte pH values (details of the procedure to derive the flat band voltage shown in SI 5). The variation of the flat band voltage as a function of electrolyte pH is shown in fig. 2 (d) to (f) for the three devices. The flat band voltage of the FinFETs shifted towards more positive values with increase in electrolyte pH due to the deprotonation of the surface hydroxyl groups at the gate oxide, followed by an increase in majority charge carriers (holes) through the transistor channel linked to the decrease of the depletion region. We obtained a shift in the flat band voltage with pH of 22 mV/pH with a standard deviation in the different devices of ± 1 mV/pH. We attribute this deviation to inhomogeneity in the sensing oxide of the different devices. The pH response of FinFETs was reproducible over time and different experimental runs on all the samples we studied. The obtained sensitivity is lower in comparison to values reported in literature ( ~ 30 to 35 mV/pH for *SiO$_2$* while the Nernst limit is ~ 59 mV/pH) [37,38]. In our case this could be due to defects introduced during the growth of the sensing oxide since we could not use a dedicated chamber for the process. The change in $\Psi_0$ was ~ 175 mV in the studied pH range in all devices, which at the working point of $V_{ref}$ = 0 is falling within the working range that was chosen in characterisation of the devices at neutral pH.

**B. Output characteristics of pH response**

The output characteristics $I_{ds}$ vs. $V_{ds}$ of the three FinFETs at a fixed $V_{ref}$ = 0 V varying the electrolyte pH are shown in figs. 3 (a) to (c) with each curve corresponding to different pH values represented with the colour scale shown in (a). The $I_{ds}$ vs. $V_{ds}$ characteristics of the wires remained linear through all the electrolyte pH ranges exhibiting a good ohmic behaviour for all devices for the studied range between -50 and 50 mV. The conductance through the wire decreased with lower pH due to the repulsion of positive charge carriers by the protons adsorbed on the surface at more acidic concentrations and the subsequent increase of the depleted region. We evaluated the conductance values in this range at different electrolyte pH from the slope of the $I_{ds}$ vs. $V_{ds}$ curves.

The variation in conductance as a function of electrolyte pH at different fixed values of $V_{ref}$ from -100 mV to 100 mV in steps of 50 mV are shown in figs. 3 (d) to (f) using the same colour scale shown in the inset of fig.3 (i). An increase in conductance of the devices for decreasing $V_{ref}$ was observed because of the attracting field experienced by the positive charge carriers in the semiconductor and consecutively the larger conducting cross section of the wires. The effect of increasing $V_{ref}$ is equivalent to lowering the pH and the conductance of the devices increased with the increase of the pH of the electrolyte with a nearly linear response, which is coherent with what was observed in the transfer characteristics. We evaluated the relative change in conductance *ΔG% = ((G-G$_0$)/G$_0$)) %* for each pH step at the different $V_{ref}$ (figs. 3 (g) to (i)). *ΔG%* was higher for smaller width FinFETs, owing to the smaller dimensions resulting in an increasing impact of the depleted region on the closure/disclosure of the channel. This is in contrast to the observed behaviour in the transfer characteristics where the change in the surface potential of the oxide provided similar sensitivity for all the devices. Devices 2 and 3 showed an approximately constant behaviour of *ΔG%* reflecting a linear dependence of the output characteristics with respect to pH changes. Device 1 showed a less constant behaviour of the relative change in conductance in the investigated pH range, but with a total variation of $\Delta G\%$ of 2% in the pH range 4 to 9 it could still be considered more linear than NWs found in literature, with variation of ~ 12% in the same pH range[1,39]. As shown in fig. 1 (a) and (b) in our case with aspect ratio above 10, the depleted region ($W_d$) affects the conductance in the horizontal direction (along the width) but its effect is negligible in the vertical direction (along the height of the device the relative change ($h-W_d$)/$h$ is much smaller than the change ($W-W_d$)/$W$). These results suggests that due to the improved linearity, these devices could be measured by using their output characteristics rather than their transfer, since the small dimensions enhance the sensitivity in terms of $\Delta G\%$.

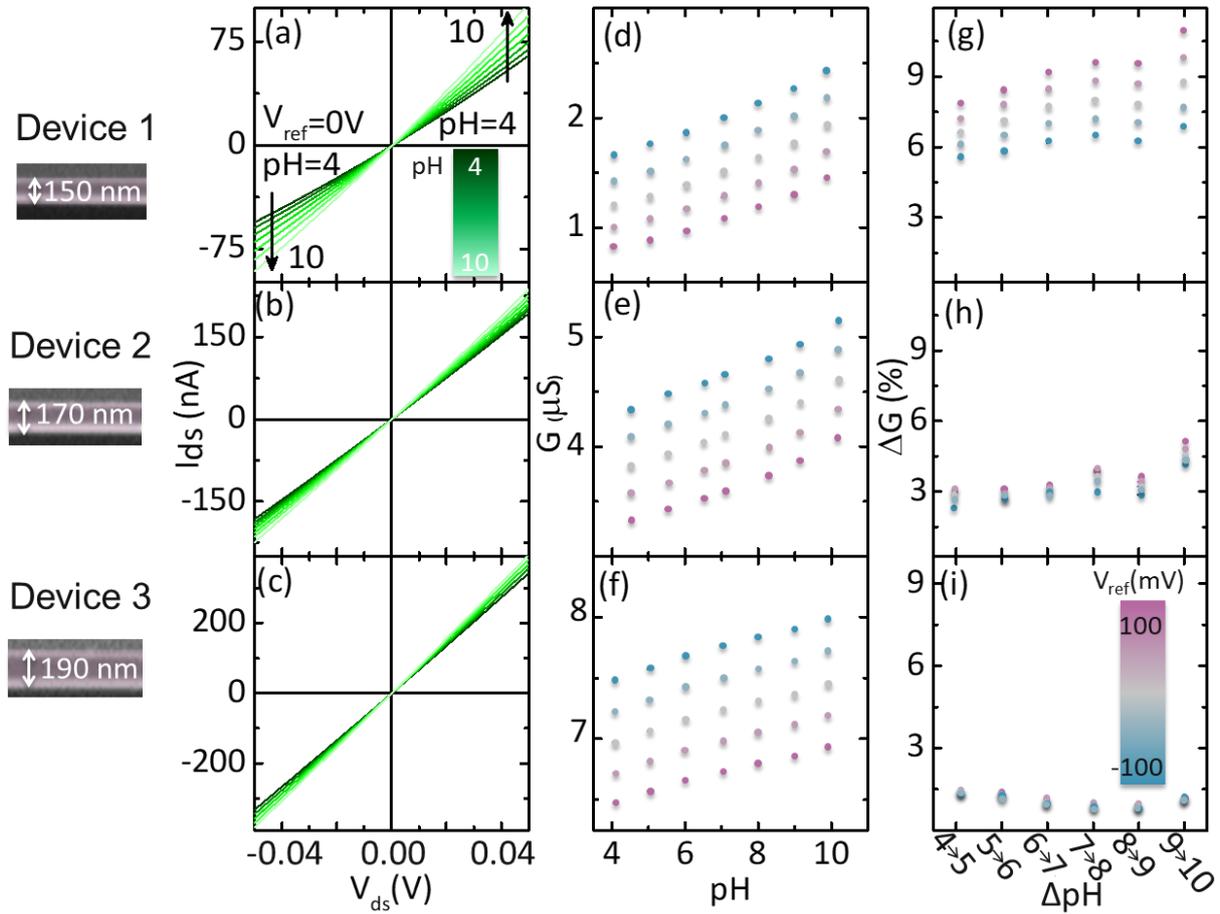

**Figure 3** Output characteristics of devices 1 to 3: (a) to (c) I-V curves at a fixed $V_{ref}$ =0 V and different electrolyte pH, from 4 to 10 in step of 1. The curves evaluated at each pH are shown using the green colour scale shown in the inset of (a); (d) to (f) conductance as a function of pH at different $V_{ref}$, varying from -100 to 100 mV in steps of 50 mV shown using the colour scale shown in (i); (g) to (i) relative conductance change percentage per unit change in electrolyte pH at different $V_{ref}$ voltages shown with the colour scale shown in (i).

## C. Theoretical and experimental data correlation

In fig. 4 (a) the conductance of the FinFET device 2 (width of 170 nm) versus the change in the surface potential is shown as black dots (showing the equivalent pH change in the top scale). We considered device 2 since its transfer characteristic allows the comparison with our Nernst-Poisson electrochemical model (details of model in SI 6) in a

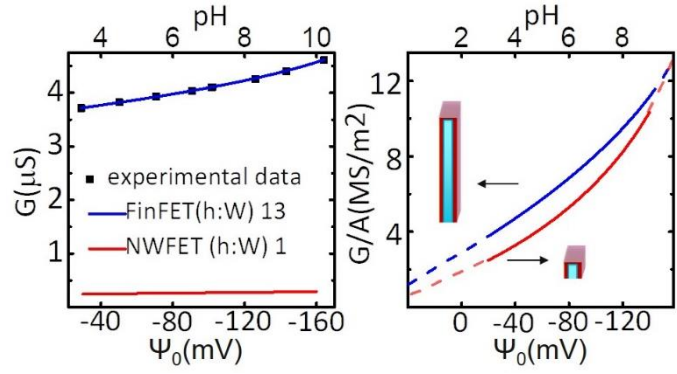

**Figure 4** (a) Device conductance vs change in surface potential: for our device the experimental points are represented with black dots and the blue line is the data fitted with our Nernst-Poisson model. In red the calculation of the conductance using the same parameters for a wire with aspect ratio 1:1 is shown. (b) Conductance normalised to the cross section area vs change in surface potential for a high aspect ratio FinFET (black) and a 1:1 aspect ratio NW (red). In both graphs the pH corresponding to the surface potential is represented on top scale.

wider range where the behaviour of the flat-band potential changed almost linearly with the pH and the dependence of $\Psi_0$ can be described with the Nernst equation[13]:

$$\frac{\partial \Psi_0}{\partial pH_B} = -2.303 \frac{kT}{q} \alpha \qquad \text{Eq. 1}$$

where $k$, $T$ and $q$ are the Boltzmann constant, absolute temperature and elementary charge, respectively. The sensitivity parameter $\alpha$ ($0 \leq \alpha < 1$) is related to the buffer capacity of the oxide that also includes other parameters like the surface charges. The Poisson equation provides the charge distribution and the width of the depleted region of the device for a $\Delta\Psi_o$ associated to a $\Delta pH$ given the doping concentration and the sensing oxide thickness that was measured by ellipsometry on a flat sample processed with the wires. We fitted the experimental data to our model shown in blue for the used FinFET with an aspect ratio of ~ 13. The parameters of the oxide thickness and channel doping obtained with the model were in good agreement with the experimental data. We obtained 20 nm for the oxide thickness both in the fitting and the experimental data obtained from the ellipsometry calibration of a probe sample used during the growth of the oxide. For the doping we obtained $2 \times 10^{17}$ cm$^{-3}$ from the fitting compared to $1\text{-}3 \times 10^{17}$ cm$^{-3}$ declared from the provider of the SOI substrate.

To analyse the response of our sensors with respect to traditional NWs, we compared the measured results of the FinFET with calculations on a device with an aspect ratio $h:W$ equal to 1:1 therefore with a square cross section. The curve calculated with the same parameters obtained for the FinFET is also shown in red in figure 4 (a). As expected, the response of the NW is much lower as the curve is nearly flat for the considered change in surface potential. Essentially we observed that the conductance of the FinFET is proportional to the cross-section and thus to the aspect ratio.

In figure 4 (b) we show the conductance normalized to the cross-sectional area from the simulations with aspect ratios 13 and 1 (blue and red, respectively). This is equivalent to comparing our device with the signal from a NW array of 13 wires that would reach approximately the same cross section. The curves have been extended beyond the experimental range to show the effect where the devices would be fully conductive ($W_D = 0$) and fully depleted ($2W_D = W$) with the dotted part of the graphs. Both devices have the same dynamic range because the depleting region is the same, and since they have the same width the electrical transport is closed at the same $\Psi_0$. In addition to the enhanced current shown in fig. 4 (a), in this case we note that the FinFET offers an increased linearity. The reason of the increase linearity was anticipated during the explanations of the changes in the relative conductivity. In the nanowire the depletion has the same relevance in the two physical dimensions of the wire ($W$ and $h$), meanwhile in the FinFET the depletion in the direction of the width has a bigger impact in the conducting cross section. The doping density, the dielectric constant and thickness of the oxide determine the dynamic range in which the sensor switches from fully conducting ($W_D = 0$) to fully depleted ($2W_D = W$), which establishes the link between the physical aspect ratio and the electrostatic parameters. Comparing the two devices, we can conclude that while an array of 13 NWs in parallel would provide similar current, the FinFET has an improved linearity using a smaller footprint, without any sacrifice of the sensitivity of the FinFET in an equal pH range.

**D. Drift and response time**

We studied the current versus time characteristics of device 2 (width of 170 nm) implementing a real-time drain current measurement at fixed $V_{ds}$ (200 mV) and $V_{ref}$ (0 V). At a fixed pH 3 $I_{ds}$ was stable over 3000 seconds time with observed drift lower than 1 nA (ca.1.5%) (data shown in SI 7). The pH of the electrolyte was also changed in a range from ~3 to ~11 in multiple cycles to study the hysteresis. We observed a change between cycles with maximum value of 15 nA evaluated at pH 3, which is similar to other values reported for $SiO_2$ pH ISFETs[38,40].

Additionally, we carried out pH measurements in which the sensor was swapped between solutions with different pH values. During these experiments, the temperature of the electrolytes was kept close to 0 °C using an ice bath to slow down the diffusion of ions in solution. The FinFET response was reproducible when the pH of the electrolyte was swapped several times from the solution at 10.5 to lower pH solutions in a range from 9.4 to 3.1 (Fig (5 (a)). During these measurements, the current of the sensor reached equilibrium after a settling time lasting several seconds, which was also reported in other pH sensing FETs in literature [41,42]. Figure 5 (b) shows the calculated settling time as function of the higher final proton concentration when the sensor was swapped from the electrolyte at pH 10.5 to the other lower pH values obtained, fitting the time response with an exponential function with a characteristic time constant (calculations details shown in SI 8). To discard the origin of the settling time from a capacitive effect, we measured the real-time response by changing the reference voltage. Specifically, the reference electrode voltage was increased from 0 to 24, 96 and 165 mV in multiple cycles, which would be equivalent to the change in surface potential produced by pH changes. Figure 5 (c) reports $I_{ds}$ vs. time at a fixed $V_{ds}$ and pH value. $I_{ds}$ was stable within each $V_{ref}$ value and no settling time was observed. We also discarded a

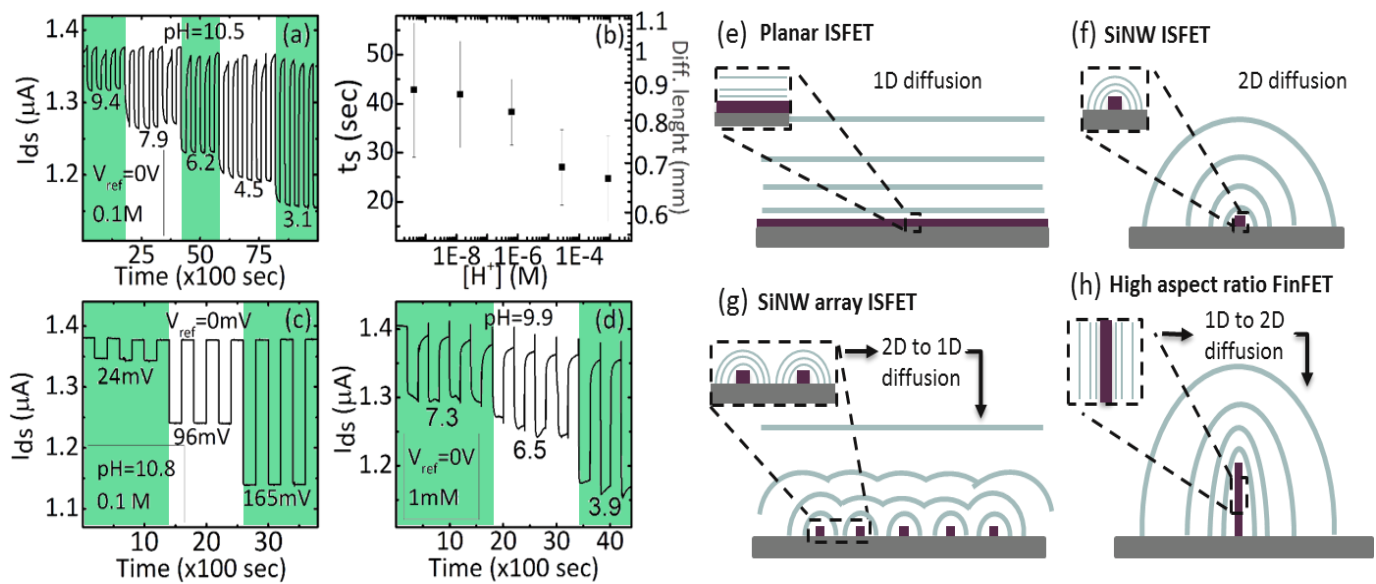

**Figure 5** (a) Real time measurment of $I_{ds}$ changing abruptly the pH from low proton concentration (pH 10.5) to higher proton concentration (changing succesively to pH 9.4, 7.9, 6.2,4.5,3.1) without rinsing. Ionic concentration during the experiment was 0.1 M. (b) Result of the corresponding settling time and diffusion length derived from fitting (a) to a characteristic setling time (details of calculations in SI 8). The large error bars correspond to the experimental dispersion orignated in the difficulty of moving the sensor between different buffers without creating turbulences in the system. (c) Real time measurment of $I_{ds}$ changing abruptly $V_{ref}$ 24, 96 and 165 (mV) equivalent to changes in the surface potential produce by the pH changes of 1, 4, and 7 at ionic concentration 0.1 M. (d) Real time measurment of $I_{ds}$ changing abruptly the pH from low proton concentration (pH 9.9) to higher proton concentration (changing succesively to pH 7.3, 6.5 and 3.9) without rinsing. Ionic concentration during the experiment was 1 M. (e) to (h) Schematic raprensentation of the diffusion lines of equal concentration for an ISFET, SiNW, SiNWs array and FinFET respectively. The sensor is represented in purple, the substrate in grey and the diffusion lines of equal concentration in light blue.

contribution to the settling time from the ionic strength due to charge screening of the ions in solution or due to the double layer capacitance by measuring the real-time response at a different electrolyte concentration. Figure 5 (d) shows $I_{ds}$ vs. time when the sensor was moved several times from the solution at pH 10.5 to the lower pH solutions at an ionic strength 1mM, two orders of magnitude lower than the experiments in fig. 5 (a). The obtained real-time pH response at two different electrolyte concentrations was similar and therefore we concluded that the change in electrolyte concentration has a negligible influence [43].

We interpreted the settling time as an effect of the proton diffusion. As described in the introduction, the analytes adsorbed by the sensing substrate create a concentration gradient in the solution. In our case when the FinFET is introduced from the higher pH solution into the lower ones the protons adsorbed create a gradient. The protons diffuse towards the oxide sensing layer until the surface saturates and the equilibrium is reached with the observed steady electrical signal. The efficiency of a sensor will depend on the sensitivity, but also on the time needed to reach the equilibrium or a significant signal for sensing. Figure 5 (e) to (h) represent the different gradient geometries created by a traditional planar ISFET, a NW, a NW array and our FinFET, respectively, showing lines of equal concentration. In a traditional planar ISFET

(fig 5 (e)) these lines are parallel to the surface of the sensor, and the diffusion occurs in one dimension, perpendicular to it. In a NW (fig. 5 (f)) the created lines of equal concentration are parallel to the axis of the wire, thus the diffusion occurs in the two perpendicular dimensions as function of the higher final proton concentration, which has been shown to provide a much shorter settling time than the case of the planar sensor[7]. NW arrays have a mixed behaviour shown in fig. 5 (g). At high concentrations the equilibrium is reached relatively faster and the particles provide mostly from a distance close to the sensor, shown in the area zoomed close to the NWs. In this case, the equal concentration lines are similar to the ones shown from the single wire in fig 5 (f). However, as the concentration is lower the molecules have to diffuse from further regions. At some point, the equal concentration lines become similar to the one of the planar sensor; this effect was described as dimensionally frustrated diffusion and decreases the overall efficiency of NW arrays because it extends the settling times to unpractical limits [35].

In accordance with Brownian motion, we calculated the associated length ($L_D$) to the diffusion time ($\tau$) of our sensors $L_D = \sqrt{2D\tau}$, using for the diffusion constant of protons in water $D$ = 9·10$^{-9}$ m$^2$/sec. We included the equivalent length scale of the diffusion time in fig. 5 (b). The further distance from which the protons diffuse towards the FinFET surface varies from half to one mm, which is three orders of magnitude larger than the height of the sensor. The reason for the small change of settling time observed in the FinFET (comparable to the standard deviation) is a consequence of the fast diffusion of protons. Other potential analytes of interest like DNA or proteins with slower diffusion constants and potential incubation times of hours, can have potentially a behaviour in which the diffusion occurs differently depending on the initial concentration. When the initial concentrations is high, the analytes can reach the sensor from close to the surface, diffusing on planar concentration fronts parallel to the surface of the sensor with a double gating effect. At lower concentrations associated with long incubation times the analytes that will reach the sensor originate from further regions and the diffusion will switch to 2D (hemicylindrical) (fig. 5 (h)). Opposite to NW arrays, the FinFET changes from a traditional ISFET like behaviour to a more efficient diffusion in 2D, which would provide an advantage to measure slow diffusing molecules at low concentrations.

## Conclusions

We have presented a rational design of a novel FinFET device for electrochemical bio sensing in which the width $W$ is connected to twice the depleted region in the dynamic range of the biosensor and has an aspect ratio $h:W$ of at least 10. We fabricated different versions of these FinFETs with different aspect ratios varying the $W$. In this article we reported three representative devices. The developed FinFETs were characterized in liquid gate configuration and showed reproducible pH sensitivity on different devices and experimental runs. The obtained pH sensitivity response in terms of variations of $V_{ref}$ to obtain the flat band potential among the different FinFETs (~ 22 ± 1 mV/pH) was low compared to the values reported earlier for silicon oxide sensing layer, but it was reproducible over different experimental runs and independent on device sizes. The pH sensitivity in terms of normalized variation of the conductance showed a dependence on the FinFETs width, increasing with decreasing in device size. This size dependent effect is inherent to the configuration of the developed FinFET sensor and was correlated with theoretical simulations. In addition, we characterised the device time response, observing a dependence of the settling time attributed to the diffusion of protons.

The FinFETs configuration that we presented shows a higher signal and improved linearity compared to Si NWs, and better linearity and lesser footprint than NW arrays prepared to have similar currents. Owing to the planar configuration of the conduction channel, in our device, the influence of small defects is localised and negligible on the sensor signal compared to their effects in small SiNWs. The developed high-aspect ratio FinFET-based sensors offer larger total surface area, which is advantageous for relatively homogeneous functionalization of required bio-chemical molecules in contrast to single NWs. We observed a dependence of the response time on the concentration of protons that we attributed to the diffusion. We argue that the proposed configuration of the device would be beneficial for applications where slow diffusing molecules are in low concentrations because it provides high signals while the geometry would benefit of the diffusion in several dimensions towards the sensor.

All these characteristics provide our sensor with promising advantages for improving sensing towards label free electrochemical sensing of charged biomarkers. We have set these advantages as a relation between the aspect ratio of the device and the depleted region during the dynamic range attributed to the analyte charges. Bio-functionalisation of the FinFETs is expected to provide soon quantification for all these potentials in the near future.

## Conflicts of interest

There are no conflicts to declare.

## Acknowledgements

This project was financed by the FNR under the Attract program, fellowship number 5718158 NANOpH. We would like to thank Sivashankar Krishnamoorthy for useful discussions and help during the project.

‡ Footnotes relating to the main text should appear here. These might include comments relevant to but not central to the matter under discussion, limited experimental and spectral data, and crystallographic data.
§ §§  etc.

44